\documentstyle[prl,epsfig,aps,amsmath,amssymb,multicol]{revtex}

\begin{document}

\draft

\title{Drain Voltage Scaling in Carbon Nanotube Transistors}

\author{M.~Radosavljevi\'c$^\ddagger$, S.~Heinze$^\ddagger$, J.~Tersoff$^*$, and Ph.~Avouris$^+$}

\address{IBM Research Division, T.~J.~Watson Research Center, Yorktown Heights, New York 10598}

\date{\today}

\maketitle

\begin{abstract}
While decreasing the oxide thickness in carbon nanotube
field-effect transistors (CNFETs) improves the turn-on behavior,
we demonstrate that this also requires scaling the range of the
drain voltage. This scaling is needed to avoid an exponential
increase in Off-current with drain voltage, due to modulation of
the Schottky barriers at both the source and drain contact. We
illustrate this with results for bottom-gated ambipolar CNFETs
with oxides of 2 and $5\,$nm, and give an explicit scaling rule
for
the drain voltage. Above the drain voltage limit, the Off-current
becomes large and has equal electron and hole contributions. This
allows the recently reported light emission from appropriately
biased CNFETs.
\end{abstract}


\vspace*{-.8cm}

\begin{multicols}{2}

\narrowtext

%
%

Recently, it has been shown
experimentally~\cite{Martel01,Appenzeller02} as well as
theoretically~\cite{Heinze02,Nakanishi02} that typical carbon
nanotube field-effect transistors (CNFETs)~\cite{Tans98-Martel98}
operate by changing the transmissivity of the Schottky barrier
(SB) at the contact between the metal electrode and the nanotube
(NT). Due to the sharp edge at the metal-NT contact and the quasi
1D-channel of the NT, this barrier can be thinned sufficiently
with gate voltage to allow thermally-assisted tunneling of
electrons or holes. This working principle has important
implications for device performance, e.g.\  it leads to a
different scaling of the turn-on gate voltage with oxide thickness
than for conventional
transistors.\cite{Appenzeller02,stefan-submitted}

Typically, CNFETs have had thick gate oxides, requiring large gate
voltage for turn-on, and could be operated at drain voltages of
$1\,$V or more. The performance of CNFETs can be systematically
improved by reducing the thickness of the gate
oxide.\cite{Appenzeller02,stefan-submitted,Bachtold01,Javey02}
However, we find that this places new constraints on the drain
voltage.  If the drain voltage is not scaled appropriately,
reducing the oxide thickness can lead to minority-carrier
injection at the drain, giving unacceptable Off-currents. The
simultaneous electron and hole currents in this regime are of
great interest for light-emission devices,\cite{misewich} which
have requirements opposite to SB-CNFETs. We derive explicit
requirements for drain voltage scaling to obtain optimal device
performance.

Our back-gated CNFET devices are produced using a two-step
oxidation process which allows the formation of small ultra-thin
oxide areas on which NTs are contacted, while keeping most of the
substrate covered with thicker field oxide.  The fabrication
begins with $100\,$nm thick silicon nitride on top of a degenerately
doped silicon wafer which serves as the backgate.  Electron beam
lithography (EBL) and reactive ion etching are used to pattern and
remove the nitride except in micrometer-sized areas.  The exposed
silicon is covered by $120\,$nm thick, thermally grown field
oxide.  Silicon nitride is dissolved in phosphoric acid, and
precisely controlled dry oxidation is used to deposit another
$2\,$nm or $5\,$nm of silicon dioxide. Carbon
nanotubes\cite{thess} are dispersed on these substrates, and
source and drain connections to NTs located in areas of thin oxide
are patterned using standard EBL and lift-off.

\begin{figure}
\begin{center}
\epsfig{file=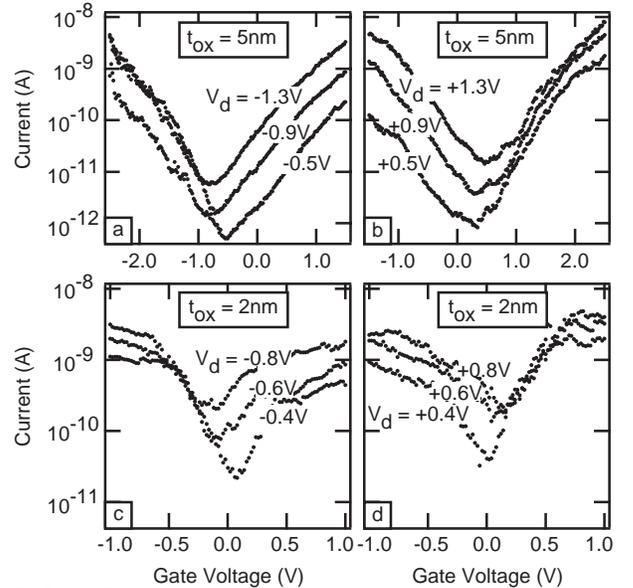,width=8.0cm,angle=0}
\caption{\label{Fig:Midgap_Experiment} Experimental $I$
vs.~$V_{\rm g}$ characteristics for scaled ambipolar CNFETs at
different $V_{\rm d}$. The device parameters are: $t_{\rm ox} =
5\,$nm and $L = 300\,$nm in (a) and (b), and $t_{\rm ox} = 2\,$nm
and $L = 60\,$nm in (c) and (d).
($L$ is the length of the NT between source and drain.) All
measurements are performed in $10^{-6}\,$Torr vacuum.}
\end{center}
\end{figure}

We focus on ambipolar devices, in particular those where the Fermi
level at the metal-NT contact falls in the middle of the bandgap.
Figure \ref{Fig:Midgap_Experiment} shows typical transfer
characteristics of current ($I$) vs.\  applied gate voltage
($V_{\rm g}$) for different drain voltages ($V_{\rm d}$) and oxide
thicknesses ($t_{\rm ox}$). All curves show the typical ambipolar
characteristics, with hole current to the left of the minimum and
electron current to the right. The gate voltage for minimum
current shifts noticeably with drain voltage. Furthermore, we
observe that at positive $V_{\rm d}$ the curves almost overlap for
the electron current while they split for the hole current, and
vice versa for negative $V_{\rm d}$.

The minimum current as a function of $V_{\rm g}$, i.e.~the
Off-current
($I^{\rm Off}$), rises steeply with increased absolute value of
$V_{\rm d}$. While the Off-current increase is clear in CNFETs
with $t_{\rm ox} = 5\,$nm, it is particularly worrisome for
devices with $t_{\rm ox} = 2\,$nm where performance deteriorates
already at $V_{\rm d} \sim 0.6\,$V. (We note for comparison that
the concurrently measured current leakage between the gate and
source/drain electrodes are below $100\,$fA and $1\,$pA for CNFETs
with $t_{\rm ox}=5\,$nm and $2\,$nm, respectively.)

%
%

\begin{figure}
\begin{center}
\epsfig{file=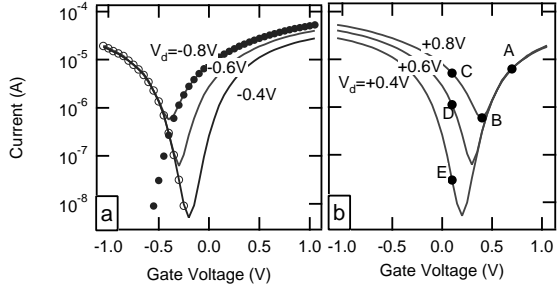,width=8.0cm,angle=0}
\caption{\label{Fig:Midgap_Theory} Effect of the drain field on
the calculated $I$ vs.\ $V_{\rm g}$ characteristics of a CNFET.
(a) and (b) curves differ only by opposite sign of $V_{\rm d}$. In
(a), open and filled circles show current decomposed into hole and
electron contributions, respectively, for $V_{\rm d} = -0.8\,$V.
Labeled points in (b) are discussed in text. The metal Fermi level
is assumed to fall at midgap in the NT at the contact. The
simulation parameters are: $t_{\rm ox} = 2\,$nm, $L = 200\,$nm,
the top gate is $120\,$nm from the NT, the contact thickness is
$20\,$nm, and the contact has length $100\,$nm. }
\end{center}
\end{figure}

In order to understand these observations we use a semiclassical
approach to calculate the total current through the device. For
ballistic transport in the NT,\cite{Fuhrer02} we can calculate the
current with the Landauer-B\"uttiker formula:
\begin{equation}
I=\frac{4e}{h} \int \left[ F(E)-F(E-eV_{\rm d}) \right] T(E) dE ~,
\label{Eq:LB_formula}
\end{equation}
where $F(E)$ is the Fermi function. The energy-dependent
transmission $T(E)$ through the device is controlled by the SBs at
the source and drain contact. The transmission through the SBs are
evaluated within the WKB approximation, using the idealized band
structure~\cite{mintmire98} for a NT with E$_{\rm g} = 0.6\,$eV
and diameter of $1.4\,$nm.

The shape of the SBs at the source and drain contact depend on the
electrostatic potential along the NT. We calculate the potential
by solving the Laplace equation for a device geometry with a
bottom gate at $t_{\rm ox}$ from the NT and a grounded top
electrode which is far from the NT. Neglecting charge on the NT is
a good approximation for the Off-state and the turn-on
regime.\cite{Heinze02,stefan-submitted}

Figure~\ref{Fig:Midgap_Theory} shows the calculated transfer
characteristics at various drain voltages, for a device geometry
as in the experiment. The effect of drain voltage on the current
characteristics for ideally ambipolar devices is clearly captured
by the model.  We see the splitting of the curves for the electron
branch at negative $V_{\rm d}$ and for the hole branch at positive
$V_{\rm d}$ as in the experiment. Moreover, in the model, the gate
voltage giving minimum current is exactly half the applied drain
voltage, and the current is symmetric about this gate voltage
$V_{\rm g} = V_{\rm d}/2$. The shift of the current minimum in the
experiment, Fig.~\ref{Fig:Midgap_Experiment}, is in approximate
agreement with this prediction. (The quantitative current is
sensitive to details of the contact geometry, and the sharper
turn-on in the theory may result from an assumed geometry more
favorable than the actual one.\cite{stefan-submitted})

%
%

\begin{figure}
\begin{center}
\epsfig{file=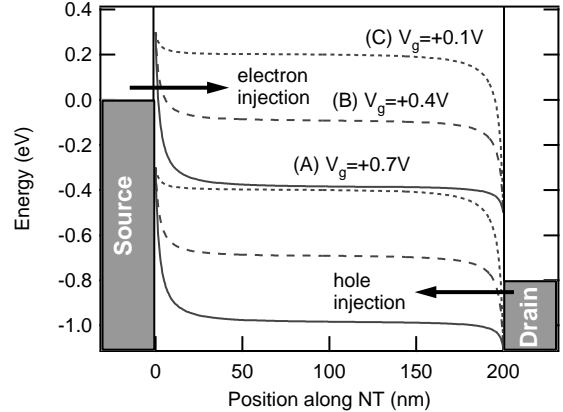,width=8.cm,angle=0}
\caption{ \label{Fig:Band-Diagrams} Band diagrams for the
simulations shown in Fig.~\ref{Fig:Midgap_Theory}, with $V_{\rm d}
= +0.8\,$V and $V_{\rm g} = +0.7\,$V (solid line), $+0.4\,$V
(dashed), and $+0.1\,$V (dotted), corresponding to points labelled
A, B, C respectively.}
\end{center}
\end{figure}

Calculated band diagrams for different voltage regimes are
displayed in Fig.~\ref{Fig:Band-Diagrams}. In the long channel
limit ($L/t_{\rm ox} \gg 1$) the electric field, as well as the
transmissivity of the source and drain contacts, should only
depend on the two potential differences $V_{\rm g}$ and $V_{\rm
d}-V_{\rm g}$. In a sense, the device acts as two
SBs connected by a field-free region in the bulk of the nanotube.
At large positive $V_{\rm g}$, labeled by point A in
Fig.~\ref{Fig:Midgap_Theory}(b) and the solid line in
Fig.~\ref{Fig:Band-Diagrams}, the CNFET is in the On-state.
Because $V_{\rm g} \sim V_{\rm d}$ at point A, the band bending at
the drain is insignificant, and the current is controlled by the
source barrier.\cite{Heinze02,Nakanishi02} As the gate voltage is
decreased, the SB at the source becomes wider and thus the
electron tunneling current is lowered. However, the potential drop
at the drain electrode, $V_{\rm d}-V_{\rm g}$, increases. At
$V_{\rm g} = V_{\rm d}/2$ (B) both SBs have the same shape and
there is an equal contribution of electron and hole injection. If
$V_{\rm d}/2$ is close to or above the turn-on voltage, these
currents can be significant and the device cannot be turned off.
(In this regime, electron-hole recombination can lead to polarized
light emission from an individual semiconducting
nanotube.\cite{misewich})

For symmetric source and drain
contacts, as in the calculation, the electron current rises
monotonically with $V_{\rm g}$ exactly as the hole current
increases monotonically with $V_{\rm d}-V_{\rm g}$,
Fig.~\ref{Fig:Midgap_Theory}(a). Therefore, the total current
becomes minimal at $V_{\rm g} = V_{\rm d}/2$. As the gate voltage
is further decreased (C) the hole current at the drain dominates
the current through the device. This drain field induced current
results in a severely deteriorated Off-current.  We calculate that
for $V_{\rm d} = 1\,$V, the
%
On/Off ratio ($I^{\rm On}/I^{\rm Off}$) decreases from about
$10^5$ at $t_{\rm ox} = 50\,$nm to less than $10^2$ at $2\,$nm.

%

However, for operation of the devices as transistors an On/Off
ratio of about $10^4$ is desirable. Using an analytic model for the
scaling of the current with $t_{\rm ox}$~\cite{stefan-submitted},
we obtain an explicit rule for $V_{\rm d}$ to meet this criterion.
The upper limit is given by $V_{\rm d} = 0.3 \, \sqrt{t_{\rm ox}
[{\rm nm}]}\,$V for the simulated bottom-gate devices. The scaling
as $\sqrt{t_{\rm ox}}$ is a consequence of the same scaling for
the turn-on voltage.\cite{stefan-submitted} (For experimental
CNFETs the allowed $V_{\rm d}$ may be a little larger due to a
less favorable contact geometry, but the same scaling rule will
still hold.) For $t_{\rm ox} = 2\,$nm, we find that to have an
On/Off ratio of at least $10^4$ requires $V_{\rm d} \le 0.4\,$V.

%
%
%
%
%
%
%

\begin{figure}
\begin{center}
\epsfig{file=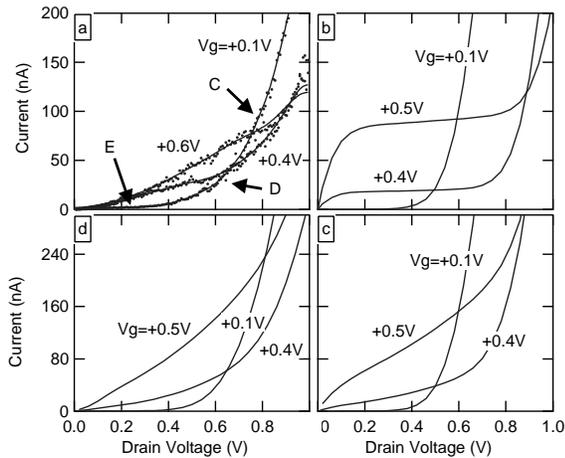,width=8cm,angle=0}
\caption{\label{Fig:output} Output characteristic for devices with
$t_{\rm ox} = 2\,$nm. (a) Experimental data for NT
with $L = 60\,$nm. Different regimes labelled C, D, and E
correspond to those in Fig.~\ref{Fig:Midgap_Theory}. (b)
Calculated output characteristic in the long channel limit. (c)
Short-channel device with symmetric geometry. (d) Short-channel
device with asymmetric geometry.}
\end{center}
\end{figure}

The drain voltage induced
minority carrier current also affects the device output
characteristics ($I$ vs.\ $V_{\rm d}$) as shown in
Fig.~\ref{Fig:output}(a). Instead of being Off near zero gate
voltage [regimes D and E in Figs.~\ref{Fig:Midgap_Theory}
and~\ref{Fig:output}(a)], a large potential difference at the
drain induces an exponential increase of the current (regime C) in
the device.  This can result in crossing of $I$ vs.\ $V_{\rm d}$
at different values of $V_{\rm g}$, a behavior not observed in
conventional transistors. The corresponding increase in
Off-current puts an upper bound on the useable range of $V_{\rm
d}$ at about $0.5\,$V, in agreement with our estimate for the
feasible On/Off ratio.

Figure~\ref{Fig:output}(b) shows that the general crossover trend
of $I$ vs.\ $V_{\rm d}$ curves at different $V_{\rm g}$ is
captured by the simulations of a long channel device which mirrors
the geometry of the measured CNFET. However, the experimental data
more closely resembles the short-channel device obtained when the
NT length is decreased
%
%
(from $60\,$nm) to $20\,$nm [Fig.~\ref{Fig:output}(c)]. Apparent
short-channel effects in our devices occur at larger $L/t_{\rm
ox}$ ratios than in the simulations, perhaps due to differences in
geometry, or to other unrecognized effects. The closest agreement
[Fig.~\ref{Fig:output}(d)] can be obtained by assuming that the
impact of the gate is slightly better at the source than the
drain, due to e.g.\  slight variations in oxide thickness or
contact geometry.
%
%
It is important to note that while these details of the geometry
affect the output characteristics, they do not change our
conclusions on the exponential increase in Off-current or the
splitting of the $I$ vs.\ $V_{\rm g}$ curves with $V_{\rm d}$.



%
%

Our discussion has been limited to a midgap line-up of the metal
Fermi energy with the NT bands, and a bandgap of $0.6\,$eV.
However, we confirmed that the drain voltage induced current is a
significant issue for ultra-thin oxide CNFETs even for asymmetric
line-ups or different band gaps. Thus the limit in drain voltage
will scale quite generally as the turn-on gate voltage, which gives
a square-root scaling for the device geometries we have studied.
\cite{stefan-submitted} Above the drain voltage limit, the device
can be used as a light source~\cite{misewich} by operating it at
the Off-voltage ($V_{\rm g} = V_{\rm d}/2$) where the current
consists of equal electron and hole contributions.

%
%

We thank J\"{o}rg Appenzeller for valuable discussions, Shalom
Wind for experimental advice, Jim Bucchignano for e-beam
lithography and Bruce Ek for expert technical assistance. The
devices were prepared with help of ASTL and CSS facilities at T.\
J.\ Watson Research Center. S.~H.~thanks the Deutsche
Forschungsgemeinschaft for financial support under Grant number
HE3292/2-1.

\bigskip

\end{multicols}
\end{document}